\documentclass[prc,twocolumn,superscriptaddress,showpacs,twoside,floatfix]%
{revtex4-1}

\usepackage{amssymb,epsfig}

\begin{document}

\title{Lifetime and fragment correlations for the two-neutron decay of $^{26}$O
ground state}

\author{L.V.~Grigorenko}
\affiliation{Flerov Laboratory of Nuclear Reactions, JINR, Dubna, RU-141980
Russia}
\affiliation{GSI Helmholtzzentrum f\"{u}r Schwerionenforschung, Planckstra{\ss}e
1, D-64291 Darmstadt, Germany}
\affiliation{National Research Center ``Kurchatov Institute'', Kurchatov sq.\ 1,
RU-123182 Moscow, Russia}
\author{I.G.~Mukha}
\affiliation{GSI Helmholtzzentrum f\"{u}r Schwerionenforschung, Planckstra{\ss}e
1, D-64291 Darmstadt, Germany}
\author{M.V.~Zhukov}
\affiliation{Fundamental Physics, Chalmers University of Technology, S-41296
G\"{o}teborg, Sweden}


\begin{abstract}
The structure and decay of $^{26}$O are investigated in a three-body
$^{24}$O+$n$+$n$ model suitable for studies of the long-lived (including
radioactivity timescale) states. We have found extremely strong effect of the
subbarrier configuration mixing on the decay width of true $2n$ emitters due to
core recoil and neutron-neutron final state interaction. This effect is far
exceeding analogous effect in the true $2p$ emitters. Our calculations provide
reasonably narrow boundaries for the lifetime vs.\ decay energy dependence for
the true $2n$ emission. An upper limit of $\sim 1$ keV for the decay energy of
the unbound $^{26}$O is inferred based on the recent experimental lifetime
value.
\end{abstract}


\maketitle


\section{Introduction}
%

True few-body decay is an exclusive quantum mechanical phenomenon comprising
simultaneous emission of several particles. Such a phenomenon leads to unusual
lifetime systematics and complex correlation patterns for decay fragments. The
term ``true'' related to a few-body decay should emphasize that the sequential
emission of particles is prohibited by specific energy conditions. These
conditions appear to be widespread beyond the nuclear driplines. Two-proton
($2p$) radioactivity \cite{Goldansky:1960} and ``democratic decay''
\cite{Bochkarev:1989} are the best-known specific examples of a broader true
few-body decay phenomenon which were under scrutinous investigation in the last
decade \cite{Pfutzner:2012}. Nowadays, the studies of the neutron-rich systems
beyond the neutron dripline are very active. Recently there is an outburst of
experimental results concerning the light true two-neutron ($2n$) emitters:
$^{10}$He \cite{Golovkov:2009,Johansson:2010,Sidorchuk:2012,Kohley:2012},
$^{13}$Li \cite{Aksyutina:2008,Kohley:2012b}, $^{16}$Be \cite{Spyrou:2012}, and
$^{26}$O \cite{Lunderberg:2012,Caesar:2012,Kohley:2013}.

\begin{figure}
\includegraphics[width=0.49\textwidth]{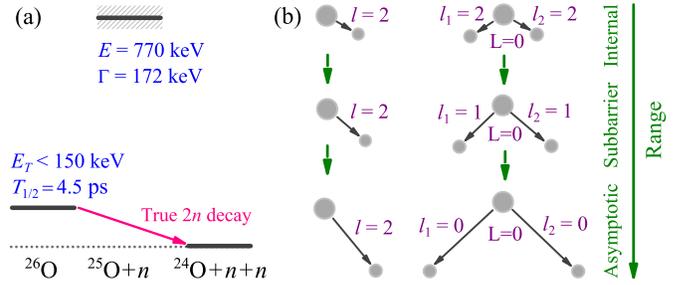}
\caption{(a) The energy scheme for true $2n$ decay of $^{26}$O. (b) Schematic
comparison of two decay mechanisms: in two-body decay the angular momentum $l$
is conserved in the process of decay (left scheme); in three-body decay of the
$[d^2]_{L=0}$ initial configuration, the subbarrier configuration mixing boosts
the penetration drastically (right scheme).}
\label{fig:mixing}
\end{figure}

The possibility of the very long-lived (radioactive) neutron emitters was
considered in our work \cite{Grigorenko:2011b}. These studies were performed
using semianalytical model based on the simplified Hamiltonians which allows the
separation of degrees of freedom. The true four-neutron emission was found to be
the most prospective phenomenon for an experimental search. However, broad
boundaries were also established in the ``lifetime vs. decay energy'' plane for
existence of radioactive true $2n$ emitters. In particular it was done for the
$^{26}$O system, information on which was absent at that moment, as one of the
candidates to be true $2n$ emitter. Since that time, the upper limits were found
experimentally for the decay energy of $^{26}$O: $E_T=150^{+50}_{-150}$ keV
\cite{Lunderberg:2012} and $E_T<120$ keV \cite{Caesar:2012}. For known $^{25}$O
ground state decay energy of $770^{+20}_{-10}$ keV \cite{Hoffman:2008}, the
$^{26}$O is now clearly ascribed to be a true $2n$ emitter, see Fig.\
\ref{fig:mixing}(a). The comprehensive overview of experimental and theoretical
studies dedicated to $^{26}$O can be found in Ref.\ \cite{Thoennessen:2012}.

Recently the halflife time of $^{26}$O was reported to be within the
radioactivity timescale: $T_{1/2}=4.5^{+1.1}_{-1.5}$(stat)$\pm 3$(sys) ps
\cite{Kohley:2013}. Taking into account experimental complexities and novelty of
the method the authors express cautious optimism about possibility of
two-neutron radioactivity observation in their work. Prospects of the discovery
of this new type of the radioactive decay call for further focused experimental
search and deeper theoretical insight, providing the guideline for such a
search. This inspired us to perform more detailed studies of the long-lived true
$2n$ emitters by example of the $^{26}$O system. Our work unveils complicated
configuration mixing under the angular momentum barriers which should be
accounted in calculations of true $2n$ emitter's widths. The important
peculiarity of such a decay mechanism for $^{26}$O with a $[d^2]$ structure is
schematically illustrated in Fig.\ \ref{fig:mixing}(b).

\section{Theoretical model}
%

\begin{figure*}
\begin{center}
\includegraphics[width=0.6\textwidth]{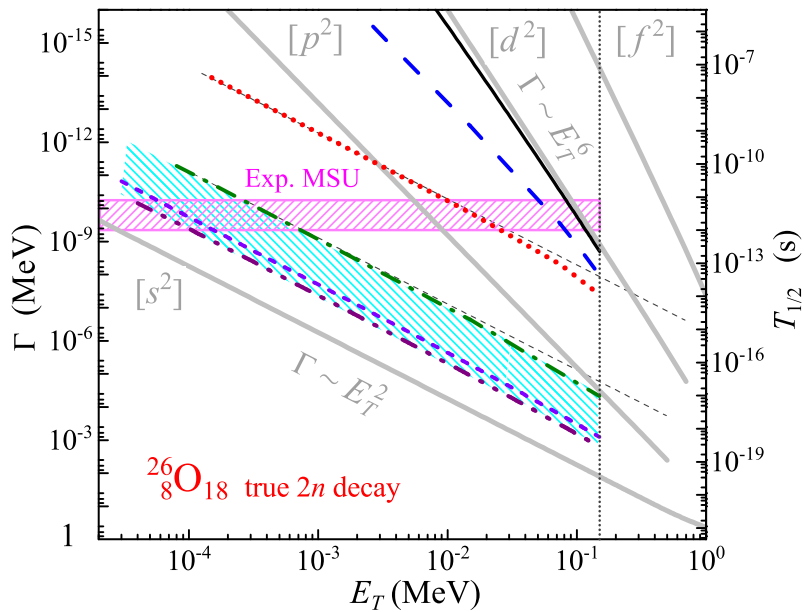}
\end{center}
\caption{(Color online) Width (halflife time) of $^{26}$O as a function of its
decay energy in different model calculations. Gray curves show the estimates of
Ref.\ \cite{Grigorenko:2011b} for decay via pure orbital $[l^2]$ configurations
coupled to the total angular momentum $L=0$. Solid curve shows the ``no FSI, no
recoil'' case, dashed curve corresponds to ``no FSI'' situation. Dotted curve
shows the results with $n$-$n$ FSI scaled by factor 0.25. Dash-dotted and
dash-double-dotted curves correspond to ``strong repulsion'' and ``moderate
repulsion'' in $s$- and $p$-waves in $^{25}$O, while short-dash curve shows the
case with ``no repulsion in $p$-wave''. Hatched areas give the experimental
limits from Refs.\ \cite{Lunderberg:2012,Kohley:2013} and the realistic
theoretical limits from this work.}
\label{fig:lifetime}
\end{figure*}

In studies of long-lived true $2n$ emitters we use the three-body
hyperspherical harmonics (HH) cluster model developed for studies of $2p$
radioactivity \cite{Grigorenko:2000}. The model demonstrated applicability and
high precision for broad range of decay energies, masses, and structures of $2p$
precursors \cite{Pfutzner:2012,Egorova:2012}. In this approach the following
sequence of problems is solved:
\begin{eqnarray}
(\hat{H}_3 - E_T)\Psi_{\text{box}} = 0 \,, \label{eq:shred1} \\
(\hat{H}_3 - E_T)\Psi^{(+)} = -i(\Gamma_{\text{arb}}/2) \Psi_{\text{box}}\,,
\label{eq:shred2} \\
\Gamma = j/N \,, \label{eq:shred3}
\label{eq:shred}
\end{eqnarray}
where $\hat{H}_3$ is three-body Hamiltonian for $^{24}$O cluster and two
neutrons. The eigenenergy $E_T$ and corresponding eigenfunction
$\Psi_{\text{box}}$ are found as a solution of the homogenous Schr\"odinger
equation (\ref{eq:shred1}) with some ``box'' boundary conditions; next, the wave
function (WF) $\Psi^{(+)}$ with a pure outgoing three-body asymptotic is found
by solving the inhomogeneous Schr\"odinger equation (\ref{eq:shred2}) with an
arbitrary value of the width $\Gamma_{\text{arb}}$; finally, the actual width
$\Gamma$ is defined in (\ref{eq:shred3}) via the outgoing flux $j$ and internal
normalization $N$ connected with WF $\Psi^{(+)}$. The width values obtained in
model calculations are shown in Figure \ref{fig:lifetime}.

\section{Three-body correlations}
%

\begin{figure}
\begin{center}
\includegraphics[width=0.47\textwidth]{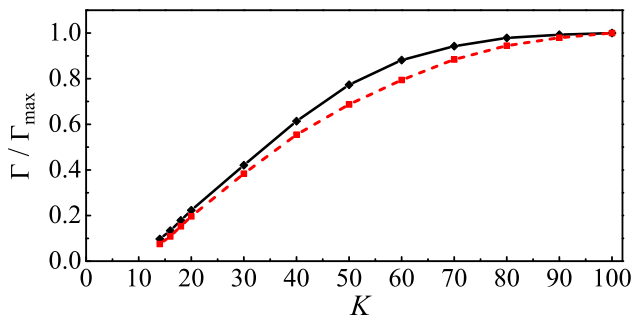}
\end{center}
\caption{Convergence of a width $\Gamma$ as a function of generalized
(hyperspherical) angular momentum $K$. Shown for the ``moderate repulsion''
case. Solid and dashed curves show calculations with $E_T=150$ and $E_T=1.5$
keV. For calculations above $K=20$ the hyperspherical channels are taken into
account adiabatically.}
\label{fig:converg}
\end{figure}

The three-body calculations in the HH method utilize the collective
coordinates: the hyperradius $\rho$ (describing collective radial motion) and
the hyperangle $\theta_{\rho}$ (responsible for geometry of the system at given
$\rho$):
\begin{eqnarray}
\mathbf{x} &=& \sqrt{\textstyle \frac{A_1A_2}{A_1+A_2}} (\mathbf{r}_1 -
\mathbf{r}_2) , \nonumber  \\
\mathbf{y}  &=& \sqrt{\textstyle \frac{(A_1+A_2)A_3}{A_1+A_2+A_3}} \left(
\textstyle \frac{A_1\mathbf{r}_1+A_2\mathbf{r}_2}{A_1+A_2} - \mathbf{r}_3
\right),
\nonumber \\
\rho &=& \sqrt{x^2+y^2} \;, \quad \theta_{\rho}=\arctan(x/y) \, .\nonumber
\label{eq:coord}
\end{eqnarray}
The boundary conditions of the HH method for the three-body decays without
Coulomb interactions are well known. However, several effects, including the
large scattering length in the $n$-$n$ channel, lead to slow radial and basis
convergence of the calculations. The hyperradii up to 2000 fm are used depending
on the decay energy. Some examples of the basis convergence are given in Figure
\ref{fig:converg}.

\begin{figure*}
\begin{center}
\includegraphics[width=1.01\textwidth]{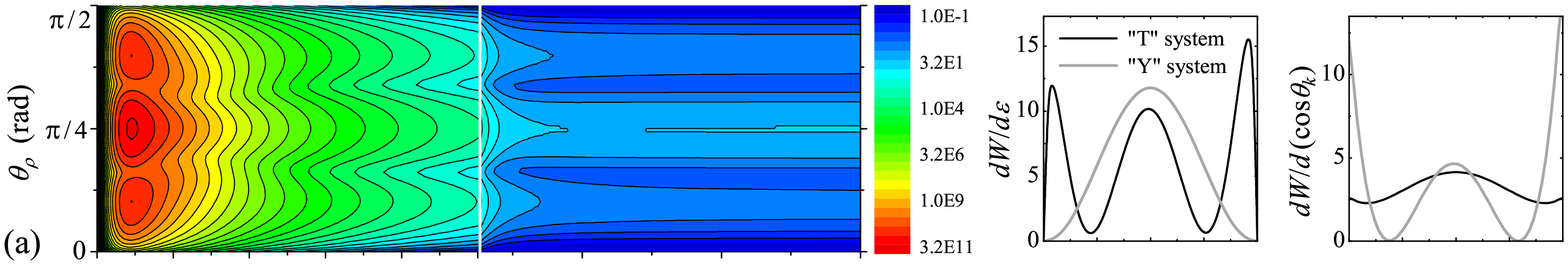}
\includegraphics[width=1.01\textwidth]{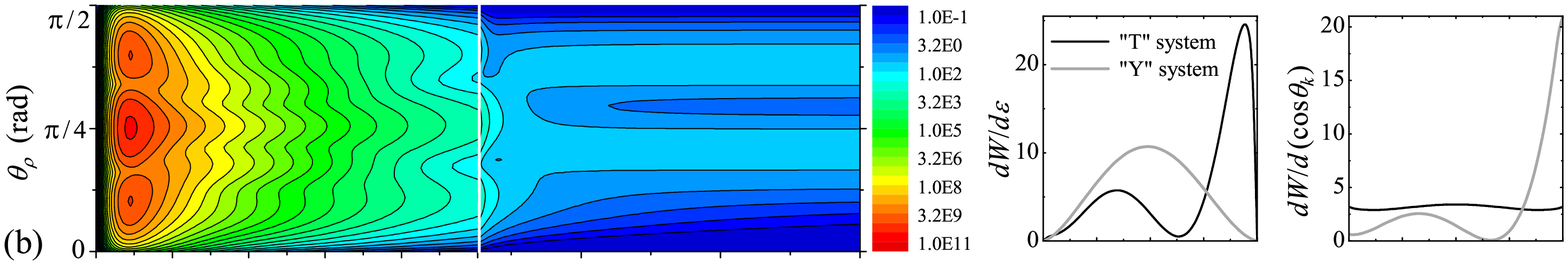}
\includegraphics[width=1.01\textwidth]{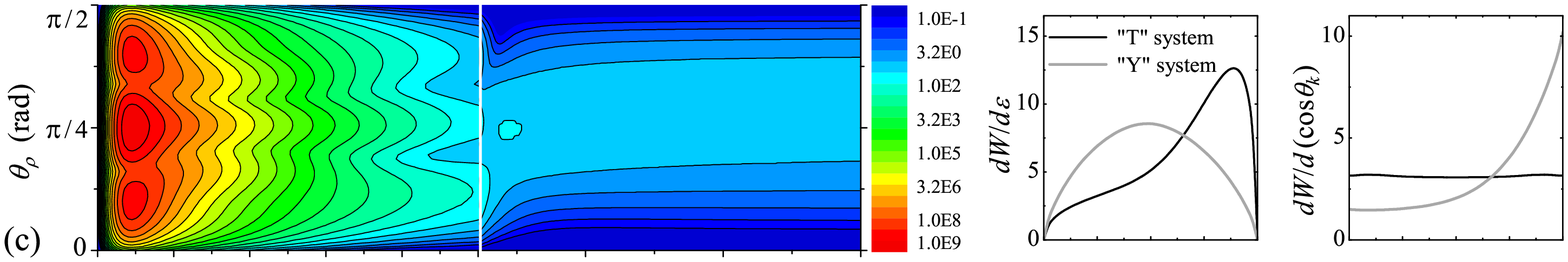}
\includegraphics[width=1.01\textwidth]{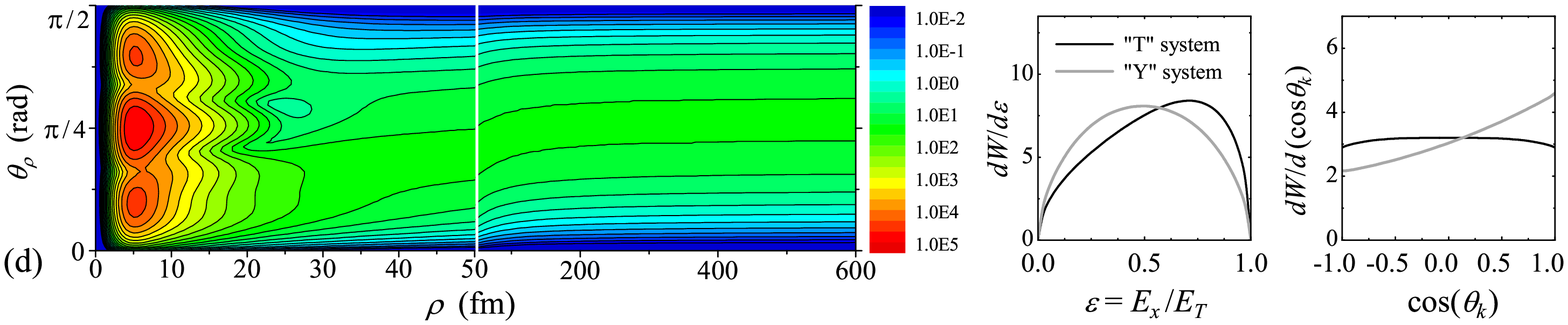}

\end{center}
\caption{(Color online) Correlations in $^{26}$O in different models: (a) ``no
FSI, no recoil'', (b) ``no FSI'', (c) ``1/4 $n$-$n$ FSI'' (d) ``moderate
repulsion''. Each row at two left panels shows the coordinate space correlation
density for $^{26}$O WF $\Psi^{(+)}$ on two different scales. Two right panels
show the energy and angular (momentum space) correlations in ``T'' and ``Y''
Jacobi systems. All cases are calculated with $E_T=75$ keV.}
\label{fig:corel}
\end{figure*}

In the momentum space the three-body energy-angular correlations (which actually
can be measured, see \cite{Pfutzner:2012}) are defined by the energy
distribution parameter $\varepsilon$ and the angle $\theta_{k}$ between Jacobi
momenta $\mathbf{k}_x$ and $\mathbf{k}_y$ (conjugated to Jacobi radii
$\mathbf{x}$ and $\mathbf{y}$):
\begin{eqnarray}
E_x = \textstyle \frac{A_1+A_2}{2 A_1A_2M} k^2_x \,  , \quad E_y = \textstyle
\frac{A_1+A_2+A_3}{2 (A_1+A_2)A_3M} k^2_y \,
, \nonumber  \\
\varepsilon = \frac{E_x}{E_x+E_y} = \frac{E_x}{E_T} \, , \quad \cos(\theta_{k})
= \frac{(\mathbf{k}_x,\mathbf{k}_y)}{k_x k_y}\, ,    \nonumber
\label{eq:mom}
\end{eqnarray}
where $M$ is an average nucleon mass. For systems consisting of two nucleons and
``core'' these correlations could be defined in two Jacobi systems: ``T''
(particle 3 is the core) and ``Y'' (particle 3 is one of the nucleons).
Coordinate and momentum space correlations for several cases of $^{26}$O
calculations are given in Figure \ref{fig:corel}.

\section{Potentials}
%

Taking into account the limited information on $^{24}$O-$n$ subsystem and
exploratory character of our current studies, we use simplified potential sets.
The $n$-$n$ potential, acting only in $s$-wave, is taken in the Gaussian form
$V(r)= V_0 \exp[-(r/r_0)^2]$, $V_{0}=-31$ MeV, $r_0=1.8$ fm \cite{BJ}. For the
$^{24}$O-$n$ channel we use a Woods-Saxon potential depending on the angular
momentum. The only available data on $^{25}$O spectrum \cite{Hoffman:2008}
allows to identify the ground state energy of $770^{+20}_{-10}$ keV and the
width of $172 \pm 30$ keV which reliably indicates a $d$-wave state. Therefore,
for the $d$-wave in the $^{24}$O-$n$ channel we use the potential with depth
$V_d=-35.5$ MeV, radius $r_0=3.5$ fm, and diffuseness $a=0.75$ fm, which well
reproduces the experimental properties of the resonance. The potential
parameters for $s$- and $p$-waves used in different model calculations are given
in Table \ref{tab:poten}. The $ls$ forces are not used in our calculations.
Their effect on the decay width is expected at the level of ten(s) of percent
(which is much smaller than other effects found in this paper). Moreover, in the
absence of the detailed experimental information on the $^{25}$O level scheme,
the use of $ls$ forces may just add an uncontrollable uncertainty to the
calculations. The three-body potential depending on $\rho$ with the typical
Woods-Saxon parameterization \cite{Pfutzner:2012} is used to control the decay
energy $E_T$ in the systematic lifetime calculations.

\section{Calculations without $n$-$n$ final state interaction (FSI)}
%

The structure of $^{26}$O is expected to be dominated by the $[d^2]$
configuration. If the decay proceeds also via the $[d^2]$ configuration then the
three-body model should provide the decay widths which have to be very close to
the $[d^2]$ estimate results from \cite{Grigorenko:2011b}. To reproduce formal
conditions of this estimate in the three-body model, two actions are required:
(i) the $n$-$n$ potential should be put to zero and (ii) the core mass should be
infinite (actually the mass number $\sim 100$ is a sufficient approximation).
One can see in Fig.\ \ref{fig:lifetime} that such three-body calculations (the
solid curve) are very close to the generic $[d^2]$ estimate, well following the
energy trend $\Gamma \sim E_T^6$ expected for the $[d^2]_0$ decay.

The correlation patterns for this case are given in Fig.\ \ref{fig:corel}(a).
One can see that the ``triple-ridge'' correlation connected with $[d^2]_0$
coupling in the internal region survives to the asymptotics producing very
symmetric ``triple-peak'' correlation in the energy distribution between
neutrons in the Jacobi ``T'' system.


\section{The recoil effect}
%

The surprise comes when the core recoil is treated correctly with the right
mass of the $^{24}$O fragment used in the ``no FSI'' calculations. The
accounting for the recoil results in several orders of magnitude larger widths
(the dashed curve in Fig.\ \ref{fig:lifetime}). The slope of the curve also
evolves with the decay energy $E_T$. At the highest energy it is parallel to the
$[d^2]_0$ curve from \cite{Grigorenko:2011b}. With energy decrease it changes
first to $[p^2]$ then to $s$-wave slope at very small $E_T$. This indicates that
the core recoil leads to configuration mixing allowing the WF to ``migrate''
into partial wave configurations with lower centrifugal barriers thus boosting
the penetration, see Fig.\ \ref{fig:mixing}(b).

Partial wave analysis and also the analysis of correlations confirm this idea.
One can see in  Fig.\ \ref{fig:corel}(b) that the triple-ridge $[d^2]_0$
internal configuration is smearing out with radius increase and at about
$\rho=150$ fm it is replaced with a double-ridge configuration typical for a
$p$-wave dominance. On asymptotics, this is reflected by a double-peak energy
correlation between neutrons.

\begin{table}[b]
\caption{Parameters of the  Woods-Saxon potential in $^{24}$O-$n$ channel for
different model cases. Radial parameters are in fm and $V_l$ are in MeV.}
\begin{ruledtabular}
\begin{tabular}[c]{ccccc}
Case  & $V_s$ & $V_p$ & $r_0$ & $a$  \\
\hline
 ``1/4 $n$-$n$ FSI'', ``Moderate repulsion'' & 70 & 70 & 3.5 & 0.75 \\
 ``no FSI'', ``Strong repulsion''    & 120 & 120 & 5 & 1.2  \\
 ``No repulsion in $p$-wave'' & 70 & 0 & 3.5 & 0.75  \\
\end{tabular}
\end{ruledtabular}
\label{tab:poten}
\end{table}


\section{Effect of the $n$-$n$ FSI}
%

This effect is found to be much stronger in the case of true $2n$ decay of
$^{26}$O than it was typically observed in the case of the $2p$ radioactivity
(true $2p$ decay). For illustration how this mechanism is engaged we performed
the ``1/4 $n$-$n$ FSI'' calculation with the $n$-$n$ potential multiplied by the
0.25 factor. Though giving only 4 times larger width than the ``no FSI''
calculation at $E_T=150$ keV, this model at smaller decay energies rapidly
sticks to the $[s^2]_0$ systematics providing dramatically larger widths. The
triple-ridge correlation in the coordinate space pattern is now dissolved at
much smaller radii [$\rho \sim 80$ fm, see Fig.\ \ref{fig:corel}(c)], compared
to the ``no FSI'' case. The energy correlation here shows one strongly
asymmetric peak indicating the $[s^2]/[p^2]$ configuration mixing, which is
consistent with the trend of the $[s^2]$ systematics.

The full $n$-$n$ FSI provides further drastic increase of the width (the
dash-double-dotted curve in Fig.\ \ref{fig:lifetime}). The energy dependence is
now completely following the $[s^2]$ trend. The coordinate correlation patterns
indicate very fast (just above $\rho \sim 30 $ fm) transition from the $[d^2]$
to the $[s^2]$ configuration, see Fig.\ \ref{fig:corel}(d). The energy
correlation between neutrons is now only slightly asymmetric indicating a
dominance of the $[s^2]$ configuration at the asymptotics already at $E_T=75$
keV.


\section{Effect of the occupied orbitals}
%

In our model we use the repulsion in $s$- and $p$-waves as a simple method
to account for the effect of the occupied orbitals in $^{24}$O in the absence of
the experimental information about these interactions. To understand the
importance of this aspect of the $^{25}$O-$n$ interaction for the $^{26}$O
lifetime, we have performed calculations with very strong repulsive interactions
in the $s$- and $p$-waves (the ``strong repulsion'' case) and without any
nuclear interaction in the $p$-wave (the ``no repulsion in $p$-wave'' case). The
widths in these calculations are different by a factor of 50 (see Fig.\
\ref{fig:lifetime}), while the lifetime energy systematics and correlations
patterns are not varying on a significant level. We may consider the band
between the latter two cases as the realistic prediction range defined by the
uncertainty of this aspect of our model. It should also be noted that the
attractive $s$-wave interaction may further increase the provided widths.
However, at the moment there is no evidence for such a possibility in the
available experimental data.


\section{Discussion}
%

Absence of Coulomb interaction in the case of long-lived $2n$ emitters,
compared to the $2p$ case, does not simplify, but makes the problem more complex
in the sense of the sensitivity to several not well studied aspects of few-body
dynamics. Nevertheless, we establish much narrower boundaries for the $^{26}$O
lifetime value thus providing the conservative upper limit of $\sim 1$ keV for
the decay energy (assuming the lifetime from \cite{Kohley:2013}).

If the experimental $^{26}$O lifetime reported in Ref.\ \cite{Kohley:2013} is
confirmed, the theoretically derived decay energy is very small in the nuclear
scale: the energy of $^{26}$O practically coincides with the $2n$ threshold.
Small separation energy of nucleon(s) is prerequisite for formation of halo
structures. Formally, the radius of the decaying system (negative separation
energy) is infinite. Practically, for systems with radioactive lifetimes, the
radial characteristics are reliably saturated for integration in the subbarrier
region and we can investigate long-lived $^{26}$O in terms of halo structure. In
our calculations we have found values around 5.7 fm for the ``valence'' neutron
rms radius in $^{26}$O. Such values are typical for $^{11}$Li which possesses
the most extreme $2n$ halo known so far. The huge halo of $^{11}$Li is connected
mainly with the important $[s^2]$ component of the WF. In $^{26}$O the analogous
radial properties are predicted due to rise of the $[s^2]$ component on the
asymptotics, despite the $[d^2]$ component strongly dominates in the nuclear
interior. The possible halo structure of the heavy oxygen isotopes was discussed
in Ref.\ \cite{Ren:1995}, where $^{26}$O was predicted to be bound. Interesting
to note that the maximum radial extent of the valence neutron configuration
(around 4.4 fm) predicted in this work is much smaller compared to our results,
which are based on a very careful treatment of the asymptotic part of the WF.

The decay energy of less than 1 keV expected from experimental lifetime
underlines the importance of special experimental techniques aimed at such
phenomena. The direct measurement of the $^{26}$O decay energy can provide an
important cross check for possible discovery of the $2n$ radioactivity in
\cite{Kohley:2013}. However, such small decay energies are not accessible for
existing experimental setups: what we know so far is just the fact that $^{26}$O
decays via $2n$ emission with intractably small decay energy. A method, based on
the precision measurements of neutron angular correlations in reactions with
relativistic secondary beams, could overcome this problem. This method was
proposed in Ref.\ \cite{Grigorenko:2011b} for future searches of $2n$ and $4n$
radioactivities and it is naturally suited for extreme low decay energies.
Conceptually analogous method for decays with proton emission was found to be
very robust \cite{Mukha:2007}.


\section{Conclusions}
%

The detailed studies of possible long-lived (radioactive) true $2n$
emitters are performed for the first time by example of the $^{26}$O system. The
following main results should be emphasized.

\noindent (i) The fine few-body effects play extremely important role in the
decay dynamics of the true $2n$ emitters. The sensitivity of decay width to (a)
configuration mixing due to core recoil, (b) subbarrier configuration mixing
caused by $n$-$n$ FSI, and (c) occupied orbitals effects far exceeds the
corresponding effects in true $2p$ decays. Unexpectedly, the lifetime
systematics of $^{26}$O with $[d^2]$ internal structure sticks to the typical
$[s^2]$ behavior due to configuration mixing, see Fig.\ \ref{fig:mixing}(b).
Studies of the correlation patterns provide deep insights in the $2n$ decay
dynamics.

\noindent (ii) The performed theoretical calculations give much narrower
limits for the lifetime vs.\ decay energy dependence of true $2n$ decay of
$s$-$d$ shell nuclei than those found in Ref.\ \cite{Grigorenko:2011b}. These
narrower limits also provide much more stringent limits on the decay energies at
which the $2n$ radioactivity may be found.

\noindent (iii) We discuss the recently published evidence for a $2n$
radioactivity of $^{26}$O (its lifetime is reported to be in a picosecond
range). Being confirmed such a lifetime leads to extremely low decay energy of
$^{26}$O (below $\sim 1$  keV) and calls for special experimental techniques for
further studies of this phenomenon.

%
\textit{Acknowledgments.}
%
%
--- L.V.G.\ is supported by the Helmholtz Association under grant agreement
IK-RU-002 via FAIR-Russia Research Center and by Russian Foundation for Basic
Research 11-02-00657-a and Ministry of Education and Science NS-215.2012.2
grants.



\end{document}